\documentclass[longauth]{aa} 

\usepackage{multicol}
\usepackage{longtable}
\usepackage{natbib}

\usepackage{graphicx}
\usepackage{txfonts}

\begin{document}

   \title{The Gaia-ESO Survey: radial distribution of abundances in the Galactic disc from open clusters and young field stars}

   \author{L. Magrini\inst{1}, S. Randich\inst{1}, G. Kordopatis\inst{2}, N. Prantzos\inst{3}, D. Romano\inst{4}, A. Chieffi\inst{5},  M. Limongi\inst{6}, 
   P. Fran\c cois\inst{7}, E. Pancino\inst{1,8},  E. Friel\inst{9}, A. Bragaglia\inst{4}, G. Tautvai\v{s}ien\.{e}\inst{10}, L. Spina\inst{11}, J. Overbeek\inst{9}, T. Cantat-Gaudin\inst{12}, P. Donati\inst{4}, A. Vallenari\inst{12}, R. Sordo\inst{12},   F. M. Jim\'{e}nez-Esteban\inst{13,14}, B. Tang\inst{15}, A. Drazdauskas\inst{10},  S. Sousa\inst{16}, S. Duffau\inst{17,18}, P. Jofr\'e\inst{19,20}, G. Gilmore\inst{19}, S. Feltzing\inst{21},   E. Alfaro\inst{22},  T. Bensby\inst{21},  E. Flaccomio\inst{23}, S. Koposov\inst{19}, A. Lanzafame\inst{24}, R. Smiljanic\inst{25}, A. Bayo\inst{26,27}, G. Carraro\inst{28}, A.~R. Casey\inst{19},  M. T. Costado\inst{22}, F. Damiani\inst{23}, E. Franciosini\inst{1}, A. Hourihane\inst{19}, C. Lardo\inst{29},  J. Lewis\inst{19},  L. Monaco\inst{30}, L. Morbidelli\inst{1}, G. Sacco\inst{1}, L. Sbordone\inst{28}, C.C. Worley\inst{19},   S. Zaggia\inst{12}  }

 \institute{INAF - Osservatorio Astrofisico di Arcetri, Largo E. Fermi, 5, I-50125 Firenze, Italy
\email{laura@arcetri.astro.it} \and 
Leibniz-Institut fur Astrophysik Potsdam (AIP), An der Sternwarte 16, D-14482 Potsdam, Germany \and
Institut d'Astrophysique de Paris, 75014 Paris, France, and the Universit\'e Pierre et Marie Curie, Paris, France \and
INAF, Osservatorio Astronomico di Bologna, Via Ranzani 1, I-40127 Bologna, Italy \and
INAF, Istituto di Astrofisica e Planetologia Spaziali, Via Fosso del Cavaliere 100, I-00133 Roma, Italy\and
INAF, Osservatorio Astronomico di Roma, Via Frascati 33, I-00078 Roma, Italy\and
Paris-Meudon Observatory, 61 Avenue de l'Observatoire, F-75014 PARIS, France \and
ASI Science Data Center, via del Politecnico snc, 00133 Roma, Italy\and
Department of Astronomy, Indiana University, Bloomington, IN, USA\and
Institute of Theoretical Physics and Astronomy, Vilnius University, Saul\.{e}tekio al. 3, LT-10222 Vilnius, Lithuania\and
Universidade de Sao Paulo, IAG Departamento de Astronomia, Rua do Matao 1226, 05509-900, Sao Paulo, Brasil\and
Dipartimento di Fisica e Astronomia, Universit\'a di Padova, vicolo Osservatorio 3, 35122, Padova, Italy \and
 Centro de Astrobiolog\'{\i}a (INTA-CSIC), Departamento de Astrof\'{\i}sica, PO Box 78, E-28691, Villanueva de la Ca\~nada, Madrid, Spain\and
     Suffolk University, Madrid Campus, C/ Valle de la Vi\~na 3, 28003, Madrid, Spain\and
Departamento de Astronomia, Casilla, 160-C, Universidad de Concepcion, Concepcion, Chile\and
Instituto de Astrofisica e Ciencias do espa\c co -CAUP, Universidade do Porto, Rua das Estrelas, P-4150-762 Porto, Portugal\and
Millennium Institute of Astrophysics, Santiago, Chile\and
Pontificia Universidad Cat\`olica de Chile, Av. Vicu\~na Mackenna 4860, 782-0436 Macul, Santiago, Chile\and
Institute of Astronomy, Madingley Road, University of Cambridge, CB3 0HA, UK\and
N\'ucleo de Astronom\'{i}a, Facultad de Ingenier\'{i}a, Universidad Diego Portales, Av. Ej\'ercito 441, Santiago, Chile\and
Lund Observatory, Department of Astronomy and Theoretical Physics, Box 43, SE-221 00 Lund, Sweden\and
Instituto de Astrof\`isica de Andaluc\`ia (IAA-CSIC), Glorieta de la Astronom\`ia, E-18008 Granada, Spain \and
INAF - Osservatorio Astronomico di Palermo, Piazza del Parlamento 1, 90134, Palermo, Italy \and 
Dipartimento di Fisica e Astronomia, Sezione Astrofisica, Universit\'{a} di Catania, via S. Sofia 78, 95123, Catania, Italy\and
Nicolaus Copernicus Astronomical Center, Polish Academy of Sciences, ul. Bartycka 18, 00-716, Warsaw, Poland\and
Facultad de Ciencias, Instituto de F\`sica y Astronomia, Universidad de Valpara\`iso, Av. Gran Bretana 1111, 5030 Casilla, Valpara\`so, Chile\and
Max Planck Institut fur Astronomie, Koenigstuhl 17, D-69117 Heidelberg, Germany \and
ESO, Alonso de Cordova 3107, 19001, Santiago de Chile, Chile\and                          
Astrophysics Research Institute, Liverpool John Moores University, 146 Brownlow Hill, Liverpool L3 5RF, UK\and
Departamento de Ciencias F\`isicas, Universidad Andres Bello, Rep\`ublica 220, Santiago, Chile
}
    \date{Received ; accepted }

 
  \abstract
   {The spatial distribution of elemental abundances in the disc of our Galaxy gives insights both on its assembly process and subsequent evolution, and  on the stellar nucleogenesis of the different elements.  Gradients can be traced using several types of objects as, for instance, (young and old) stars, open clusters, HII regions, planetary nebulae. }
   {We aim at tracing the radial distributions of abundances of  elements produced through different nucleosynthetic channels  --the $\alpha$-elements   O, Mg, Si, Ca and Ti, and the iron-peak elements Fe, Cr, Ni and Sc -- by using the Gaia-ESO {\sc idr4} results of open clusters and young field stars. }
   {From the UVES spectra of member stars, we determine the average composition of clusters with ages $>$0.1~Gyr.  
   We derive statistical ages and distances of field stars. We trace the abundance gradients using the cluster and field populations and we compare them with a chemo-dynamical Galactic evolutionary model.    }
   {The adopted chemo-dynamical model, with the new generation of metallicity-dependent stellar yields for massive stars, is able to reproduce
   the observed spatial distributions of abundance ratios, in particular the  abundance ratios of [O/Fe] and [Mg/Fe] in the inner disc (5~kpc$<$R$_{\rm GC}<$7~kpc), with their differences, 
    that were usually  poorly explained by chemical evolution models. 
    }
   {Often, oxygen and magnesium are considered as equivalent in tracing 
$\alpha$-element abundances and in deducing, e.g., the formation time-scales of different Galactic stellar populations. In addition, often  [$\alpha$/Fe] is computed  combining  several $\alpha$-elements. 
   Our results indicate, as expected,  a complex and diverse nucleosynthesis of the various $\alpha$-elements, in particular in the high metallicity regimes, pointing towards a different origin of these elements
   and highlighting the risk of considering them as a single class with common features.  }

\authorrunning{Magrini, L. et al.}
\titlerunning{Abundance ratios in the Galactic thin disc }

   \maketitle
%

\section{Introduction}

The distribution of elemental abundances in the Galactic disc provides fundamental constraints to
models of galaxy formation and evolution. 
Thanks to their wide range of ages and distances and to the much higher accuracy with which these quantities can be measured in clusters with respect to field stars, 
open clusters  are considered among the best tracers of the overall Galactic metallicity distribution and  of the thin disc abundance patterns \citep[e.g.][]{friel95, friel13}.  
The advent of multi-object high-resolution spectrometers allowed us to easily obtain spectra of many stars in the same cluster.
This permitted us to confidently determine abundances in several member stars, and thus to securely relate 
the composition (including abundances of a large number of elements belonging to different nucleosynthetic channels) to a specific location and epoch in 
the history of our Galaxy.  
Among the several on-going spectroscopic surveys, the Gaia-ESO Survey \citep[GES, ][]{gilmore12,RG13},  a ESO large public survey, is providing high resolution spectra of 
different stellar population of our Galaxy using FLAMES@VLT \citep{pasquini02}, aiming at homogeneously determining stellar parameters and abundances for a large sample of stars in the field and in Galactic open clusters.
In particular, in GES the open cluster population is well-sampled and includes clusters over a large range of ages, distances, masses, and  metallicities. 
The cluster target selection will be described in Bragaglia et al. and Randich et al. (in preparation). 

Open clusters have historically been  used  to trace the spatial distribution of metallicity in the Galactic disc. 
Since the first studies \citep{janes79,PT80, janes88, frieljanes93, piatti95}, it has been found that 
the cluster population shows a significant decrease in metallicity
with increasing distance from the Galactic Centre, the so-called radial metallicity gradient. 
Complementary to the study of the overall metallicity distribution (often approximated with [Fe/H]), the 
abundance ratios of several elements, such as 
 $\alpha$, neutron-capture, iron-peak  and odd-Z elements, can provide  insightful information  
both on the star formation history in the disc and on the nucleosynthesis processes, production sites and timescales of enrichment of 
each element. 
For a complete review of the role of open clusters in tracing the Galactic abundance distribution and 
its time evolution,  we refer the reader  to \citet{friel13}.  

In this framework, the behaviour of the so-called $\alpha$-elements -- among them O,  Mg, Si, Ca, and Ti -- is of particular interest. 
These elements are indeed mainly formed through stellar nucleosynthetic processes in massive stars. 
Consequently the timescales of their recycling in the interstellar medium is much faster than, for instance, 
that of  iron, which is mainly produced in type Ia supernovae (SNIa). 
An enhancement of their abundances with respect to iron, or to other iron-peak elements,
can reveal differences  in the star formation history of different regions  of the disc \citep[e.g., ][]{yong05}. 
For example, a rapid and intense star formation in the inner disc coupled with a slower and more recent process of star formation in the outer disc, 
with a still incomplete enrichment by SNIa, might cause an increasing  [$\alpha$/Fe] in the outer part of the disc.
This is indeed predicted by several chemical evolution models \citep[e.g., ][]{m09, m15,  kubryk13, minchev14}.  
Despite of their common producers in terms of mass range, the creation of the various $\alpha$ elements is related to processes happening  
during different burning phases in the evolution of
massive stars \citep[see, e.g.,][]{PT95}: oxygen is produced during the hydrostatic burning
in the He-burning core and in the C-shell and it is expelled
during the pre-supernova phase \citep[e.g.][]{mmh}; magnesium is produced during the hydrostatic
burning in the C shell and in the explosive burning of Ne, while the other elements --Si, Ca and Ti--  come from the 
explosive burning.

The existence of such $\alpha$-enhancement in the outer disc remains under debate. 
The first studies of the abundances of the outermost disc clusters \citep{yong05, carraro04}
observed that clusters in the outskirt of the Galaxy had an enhancement in their $\alpha$-element and rapid ({\em r}) neutron capture elements (e.g., europium).    
Based on the measured enhancement in  $\alpha$- and {\em r}-element abundance ratios, \citet{yong05}  
suggested that the outer-disc open clusters were formed in a different way than the rest of the  disc, and they proposed their formation through 
a series of merger events.
On the other hand, further works have suggested that abundances of outer-disc open clusters are instead consistent 
with scaled Solar values \citep[e.g.,][]{carraro07, sestito08, pancino10, bensby11, yong12, hayden15}. 

For the same reasons for which we expect the $\alpha$-enhancement in the outer part of the Galactic disc, 
we foresee that clusters in the very inner disc might present a depletion in $\alpha$ elements over iron with respect to Solar values. 
This is particularly true for the young populations in the inner disc that trace the full chemical evolution of the inner disc characterised by 
high infall and star formation rates. 
However, observations of young populations located in the inner parts of the Galactic disc seem to contradict the expectations of chemical evolution models built in an inside-out scenario 
(see for example Fig.~9 of \citet{minchev14}).  
For example, the observations  of  `young' $\alpha$-enhanced stars \citep{chiappini15, martig15, kordo15,yong16} with [Fe/H] ranging from 
$-0.4$ to $+0.2$ dex and located towards the Galactic Centre \citep[see, e.g.][for possible explanations of their origin- in the former, young stars formed from gas survived near to the bar with a peculiar composition, 
in the latter, massive old stars accreted mass from companion]{martig15, jofre16}  
and the hints given by young inner-disc open clusters located at R$_{\rm GC}$$\leq$7~kpc \citep{m15} are difficult to reconcile with the classical inside-out scenario.  
For these two young populations, the surprising result is the higher than expected [Mg/Fe] value -- used as a proxy of [$\alpha$/M] --  for their age and location in the disc. 
They are indeed young and presumably born in  the inner disc:  
a sub-Solar value of [Mg/Fe] should be presumed because of the high infall and high star formation rates in the central part of galaxies \citep[see, e.g.,][]{minchev14}. 
Thus even just Solar [Mg/Fe] values, as found in the open cluster Be~81 \citep{m15}, are surprising if compared with what is expected from chemical evolution, i.e.  under-Solar [Mg/Fe] values (see, e.g. Fig.9 of \citet{minchev14} and Fig.8 of \citet{kubryk15b} for oxygen).

In the present paper, we make use of the UVES results of both open clusters and Milky Way field stars to investigate the radial trends of several elements, and 
compare them with the predictions of a chemical evolution model that includes radial migration \citep{kubryk15a,kubryk15b} and new generation stellar yields for massive stars (see Prantzos et al. and 
Chieffi \& Limongi in preparation).   In the following, we indicate the model adopted in the present paper as K15-improved, with the meaning described above. 
We discuss the differences among the behaviours of the various elements,  their implications on the nucleosynthesis in massive stars and SNIa and on the formation of the disc. 

The paper is structured as follows: in Section~\ref{datared} we present the data reduction and analysis and in Section~\ref{solar} we determine our Solar scale. 
In Section~\ref{sample} we describe our sample of open clusters and in Section~\ref{field} the sample of field stars. In Sections~\ref{radial} and \ref{patterns} 
we show the radial distributions of [Fe/H] and of the abundance ratios, and the abundance patterns as function of metallicity. 
In Section~\ref{model} we present a chemical evolution  whose comparison with the data is presented in Section~\ref{comparison}. 
In Section~\ref{summary} we give our summary and conclusions. 

\section{The data reduction and analysis}
\label{datared}

The UVES spectra used in the present work have been reduced and analysed by the Gaia-ESO consortium in several Working Groups (WGs).
UVES data reduction is carried out using the FLAMES-UVES
ESO public pipeline \citep{modigliani04}. The UVES data reduction process and the determination of  the radial velocities (RVs) 
are described in \citet{sacco14}. 
Different {\sc WGs} contribute to the spectral analysis of different kinds of stars and/or setups: the data discussed in the present paper have been analysed by WG11 which is in charge of 
the analysis of the UVES spectra of F-G-K spectral type stars both in the field of the Milky Way (MW) and in  intermediate-age and old
clusters and obtained with two setups, U580 and U520. 
The UVES spectra were analysed with the Gaia-ESO multiple pipelines strategy, as described in \citet{smi14}. The results of each pipeline are combined with an updated methodology (Casey et al., in prep.) to define a final set of recommended values of the atmospheric parameters.
The results of WG11 are homogenised with the results of the other WGs using several calibrators e.g., benchmark stars and open/globular clusters
selected as described in \citet{pancino16} and adopted for the homogenisation by WG15 (Hourihane et al. in preparation). 
 The final recommended stellar parameters of most of the stars included in the fourth internal data release, hereafter {\sc  idr4}, come from the combination 
of the results of many Nodes participating to the analysis with different methods, from the equivalent width to the spectral synthesis. 
In {\sc idr4}, ten Nodes were contributing to the analysis of F-G-K UVES spectra. 
The final recommended parameters of 41\% of the stars are obtained combing the results of all ten Nodes, 
21\% of nine Nodes, 14\% of eight Nodes, 9\% of seven Nodes, 6\% of six Nodes, 4\% of five Nodes, 
3\% of four Nodes, and only 2\% of three or two Nodes.
The production of the final abundances is a complex process in which 
all Nodes give for each element the abundance line by line. 
The Node abundances (line by line) are combined to produce a final homogenised abundance (per line, per star), which are in turn  combined to produce a final recommended abundance per star. Thus it is not straighford  to keep trace of the exact lines used to 
produce the final abundance in each stars. 
The full line-list (used mainly for spectral synthesis) and the ``clean'' line-list (used mainly for equivalent width analysis), together with the source and selection of the $\log~gf$, which are both experimental or theoretical,  preferring, when available, the most precise laboratory measurements, are described in \citet{heiter15} and will be available in a forthcoming paper (Heiter et al. in preparation). 
In the following analysis, we discuss abundances normalised to our internal Solar scale, thus mitigating the effect of the $\log~gf$ choice in the comparison with literature results.  

The recommended parameters and abundances used in the present work are reported in the final {\sc GESiDR4Final} catalog, which contains the observations obtained until July 2014 and which is distributed to the whole community through the  ESO portal.  

\section{Solar abundance scale}
\label{solar}
 \begin{table}
\begin{center}
\caption{{\sc idr4} Solar parameters and abundances.  }
\tiny
\begin{tabular}{llll}
\hline\hline
Sun &  T$_{\rm eff}$    &  $\log$~g  & $\xi$  \\
       &(K)  &                       & km~$s^{-1}$ \\
\hline
 & 5740$\pm$120   &4.40$\pm$0.20 & 0.90$\pm$0.10  \\
\hline \hline
Element & Sun ({\sc idr4}*) & Sun (G07) & M67 giants ({\sc idr4}) \\
\hline
{\sc F}e&   7.48$\pm$0.06 & 7.45$\pm$0.05   & 7.48$\pm$0.09(0.02)\\
{\sc O}   &  8.78$\pm$0.11  & 8.66$\pm$0.05  & 8.76$\pm$0.11(0.02)\\
{\sc M}gI & 7.65$\pm$0.12 &7.53$\pm$0.09   & 7.63$\pm$0.12(0.02) \\
{\sc S}iI   & 7.47$\pm$0.07 &7.51$\pm$0.04  & 7.48$\pm$0.07(0.03)\\
{\sc C}aI &  6.31$\pm$0.08 &6.31$\pm$0.04  & 6.31$\pm$0.08(0.02)\\
{\sc S}cII &  3.21$\pm$0.07 &3.17$\pm$0.10 &  3.21$\pm$0.07(0.01) \\
{\sc T}iI    & 4.89$\pm$0.08 &4.90$\pm$0.06 & 4.89$\pm$0.08(0.03)\\
{\sc V}I    & 3.89$\pm$0.09 &4.00$\pm$0.02 & 4.00$\pm$0.08(0.03)\\
{\sc C}rI   & 5.60$\pm$0.10 &5.64$\pm$0.10 & 5.58$\pm$0.11(0.02)\\
{\sc N}iI & 6.23$\pm$0.09 &6.23$\pm$0.04& 6.24$\pm$0.10(0.01)\\
\hline \hline
\end{tabular}
\label{tab_sun}
\end{center}
\tiny{*Average of the several measurements on the different Solar U580 spectra from the WG11 analysis.   }
\end{table}

To obtain abundances on the Solar scale, we need to define our abundance reference.  In Table~\ref{tab_sun} we show the Solar parameters (derived in a homogeneous way as the whole {\sc idr4} sample combining the results of the Node participating to the analysis) and 
three different sets of abundances.  the Solar abundances from  {\sc idr4}, the \citet{grevesse07}'s ones, and the abundances 
of giant stars in M67. 
M67 is indeed known to have the same composition as the Sun  \citep[e.g.,][]{pasquini, on14, liu16} and thus it is useful to confirm it with the GES {\sc idr4} data. 
Furthermore we aim at checking the presence of any systematic difference between the abundances obtained for dwarf and giant stars. 
We have obtained our reference Solar 
abundances from the average values of all UVES abundance determinations (from the WG11 recommended table) in the same setup used for our science observations, U580.
We have also compared with the reference Solar abundances from \citet{grevesse07} finding a very good agreement for most elements. 
However, oxygen and magnesium are both slightly higher in the  GES Sun. GES oxygen abundance is however in good agreement with the results of \citet{caffau08} and of \citet{steffen15} both based on the [OI] 630.0nm line, which is not affected by NLTE and 3D effects. 
In addition, we report the average abundances of the three member giant stars in M67 (T$_{\rm eff}$$\sim$4800-4900 and   $\log$~g$\sim$3-3.4) from the {\sc idr4} recommended table. 
We quote both the errors on the measurement (from the {\sc idr4} recommended table) and the standard deviation of the average (in parenthesis). The  very small standard deviation 
indicates a high degree of homogeneity of the cluster and high quality of the results. 

The results shown in Table~\ref{tab_sun} indicate an identical composition of the Sun and of M67 giant stars within the uncertainties, and no evident differences between abundances 
in dwarf and giant stars. 
Moreover the higher GES abundances of O and Mg than the ones of \citet{grevesse07} are confirmed also in M67 and are likely related to the choice of the atomic data and line list for these 
elements.  
In what follows, we normalise our abundances to the Solar abundances, computed as the average of several determination from the {\sc wg11} recommended table. 
These are shown in the first column of  Table~\ref{tab_sun}. 

\section{The cluster sample}
\label{sample}

We consider the sample of clusters with ages$>$0.1~Gyr whose parameters and abundances have been delivered  in {\sc idr4}. 
The sample includes several new clusters released for the first time in {\sc idr4}:  NGC2243, Berkeley25, NGC6005. NGC6633, NGC6802, NGC2516, Pismis18 and Trumpler23 and four clusters already processed in previous data releases  and discussed in previous papers:  Berkeley 81, NGC4815, Trumpler 20, and NGC6705. Detailed analysis of NGC6802 and of Trumpler23 from {\sc idr4} data are presented by Tang et al. (submitted) and by \citet{overbeek16}, respectively. 
The radial metallicity --expressed by [Fe/H]-- distribution of the inner disc clusters is discussed in \citet{jacobson16}, while the gradient traced by the very young clusters and star-formation regions is discussed by \citet{spina17}. 
Most of our sample clusters are younger than 2~Gyr, and only the two outermost clusters, NGC~2243 and Berkeley~25 are  older than 2~Gyr (see Table~\ref{tab_clu_par} for parameters and abundances of the clusters).  
The population of young and intermediate-age open clusters is extremely useful to trace the recent chemical evolution of the Galactic disc since it is not strongly  affected 
by radial migration \citep[see, e.g.,][]{minchev14}  and it is the dominant component of the young population in the disc.

In Table~\ref{tab_clu_par} we summarise the basic properties of the sample clusters --coordinates, ages, Galactocentric distances, heights above  the plane, mean radial velocities of cluster members, median metallicity and the number of cluster member stars used to compute the metallicity and the abundances. 
We  adopt, for clusters in common, the same ages and distances as in   \citet{jacobson16}.  For the two clusters not previously analysed in GES papers, we adopt distance from the Sun from literature studies, and we re-compute Galactocentric distances and heights with R$_{0}$=8~kpc. 
For each cluster we have extracted member stars using the information on their radial velocities considering as member stars  those within 1-$\sigma$ from the cluster systemic velocity
and excluding outliers in metallicity $|$[Fe/H]$_{\rm star}-<$[Fe/H]$>|>$0.1~dex, with a larger range of 0.2~dex allowed for Be81, which has more dispersed member stars in terms of metallicity. 
For each cluster, based on stars assigned as members, we have computed 
the median elemental abundances, expressed in the form 12$+\log$(X/H), which are presented in Table~\ref{tab_clu_abu}. 
The error reported on each abundance is the dispersion (computed with the robust sigma) of cluster member abundances. 
We do not report 12$+\log$(O/H) in Tr20  because of telluric contamination of the oxygen line at 6300 \AA  and 12$+\log$(O/H) in NGC2516 and NGC6633 because of high dispersion
with the robust sigma not converging. 
The stellar parameters, radial velocities and abundances of the selected cluster members are presented in Appendix A [available online], Tables~A.1 and A.2. 

\begin{table*}
\begin{center}
\caption{Cluster parameters}
\tiny
\begin{tabular}{llllllllll}
\hline \hline
  \multicolumn{1}{c}{Id} &
  \multicolumn{1}{l}{R.A.} &
  \multicolumn{1}{r}{Dec.} &
  \multicolumn{1}{l}{Age } &
  \multicolumn{1}{c}{R$_{\rm GC}$(a) } &
  \multicolumn{1}{c}{Z } &
  \multicolumn{1}{c}{rv } &
   \multicolumn{1}{c}{[Fe/H]} &
  \multicolumn{1}{c}{n. stars} &
  \multicolumn{1}{c}{Ref. Age \& Distance} \\
  \multicolumn{1}{c}{} &
  \multicolumn{2}{c}{J2000.0} &
  \multicolumn{1}{c}{(Gyr)} &
  \multicolumn{1}{c}{(kpc)} &
  \multicolumn{1}{c}{(pc)} &
  \multicolumn{1}{c}{(km s$^{-1}$)} &
  \multicolumn{1}{c}{} &
 \multicolumn{1}{c}{} &
   \multicolumn{1}{c}{} \\
\hline
  NGC2516 				&07:58:04			&	-60:45:12 & 0.12$\pm$0.04   & 7.98$\pm$0.01    & -97$\pm$4        &+23.6$\pm$1.0	& -0.06$\pm$0.05 & 13 &  \citet{sung02}\\
  NGC6705 				&18:51:05 		&     -06:16:12  & 0.30$\pm$0.05   & 6.33$\pm$0.16    & -95$\pm$10      &+34.9$\pm$1.6	& +0.12$\pm$0.05 &  15     &\citet{cantat14}\\
  NGC4815 				&12:57:59 		&     -64:57:36  & 0.57$\pm$0.07   & 6.94$\pm$0.04    & -95$\pm$6        & -29.6$\pm$0.5	& +0.00$\pm$0.04 &  3       &\citet{friel14}\\
  NGC6633 				&18:27:15			&      +06:30:30  & 0.63$\pm$0.10   & 7.71$\pm$0.01    &  +52$\pm$2 	      &	 -28.8$\pm$1.5 & -0.06$\pm$0.06 & 8       &\citet{jeffries02}\\
  NGC6802 				&19:30:35			&      +20:15:42  & 1.00$\pm$0.10   & 6.96$\pm$0.07    &  +36$\pm$3 	      &	+11.9$\pm$0.9  & +0.10$\pm$0.02 & 8         &\citet{jacobson16}\\
  Be81 					&19:01:36        		&      -00:31:00 & 0.86$\pm$0.10  & 5.49$\pm$0.10     & -126$\pm$7      &	-126$\pm$7       & +0.22$\pm$0.07 & 13        &\citet{m15}\\
  Tr23 					&16:00:50			&      -53:31:23 & 0.80$\pm$0.10  & 6.25$\pm$0.15     & -18$\pm$2        &	-61.3$\pm$0.9  &+0.14$\pm$0.03 & 11         &\citet{jacobson16}\\
  NGC6005 				&15:55:48			&      -57:26:12 & 1.20$\pm$0.30  & 5.97$\pm$0.34     & -141$\pm$26    & -24.1$\pm$1.34 & +0.16$\pm$0.02&  7 		&\citet{piatti98} \\
  Pis18 					& 13:36:55	         &      -62:05:36& 1.20$\pm$0.40  & 6.85$\pm$0.17      & +12$\pm$2     & -27.5$\pm$0.7    & +0.10$\pm$0.01& 3  	&\citet{piatti98}\\
  Tr20 					&12:39:32 		&     -60:37:36  & 1.50$\pm$0.15 & 6.86$\pm$0.01       & +136$\pm$4      & -40.2$\pm$1.3	& +0.12$\pm$0.04  & 27   & \citet{donati14GES}\\
  Be44                                 	&19:17:12	        		&    +19:33:00  & 1.6$\pm$0.3      & 6.91$\pm$0.12       & +128$\pm$17    & -8.7$\pm$0.7 	& +0.20$\pm$0.06  & 4 &\citet{jacobson16}\\
  Be25 					& 06:41:16		&   -16:29:12    & 4.0$\pm$0.5     & 17.6$\pm$1          & -1900$\pm$200 & +136.0$\pm$0.8 & -0.25$\pm$0.05 &   6  &\citet{carraro05} \\
  NGC2243                                & 06:29:34		&    -31:17:00   & 4.0$\pm$1.2     &  10.4$\pm$0.2         & -1200$\pm$100 & +60.2$\pm$0.5 & -0.38$\pm$0.04  & 16 &\citet{BT06}\\
\hline \hline
\end{tabular}
\label{tab_clu_par}
\end{center}
\end{table*}

\begin{table*}
\begin{center}
\caption{Clusters' elemental abundances expressed in the form 12$+\log$(X/H). }
\tiny
\begin{tabular}{lrrrrrrrrr}
\hline \hline
  \multicolumn{1}{c}{Id} &
  \multicolumn{1}{c}{OI/H} &
  \multicolumn{1}{c}{MgI/H} &
  \multicolumn{1}{c}{SiI/H} &
  \multicolumn{1}{c}{CaI/H} &
  \multicolumn{1}{c}{TiI/H} &
  \multicolumn{1}{c}{ScII/H} &
  \multicolumn{1}{c}{VI/H} &
  \multicolumn{1}{c}{CrI/H} &
  \multicolumn{1}{c}{NiI/H} \\
\hline 
  NGC2516 & -                        & 7.62$\pm$0.05 & 7.34$\pm$0.07 & 6.29$\pm$0.03 & 4.96$\pm$0.08 & 3.07$\pm$0.06 & 3.99$\pm$0.06 & 5.61$\pm$0.08 & 6.13$\pm$0.04\\
  NGC6705 & 8.75$\pm$0.06 & 7.85$\pm$0.05 & 7.59$\pm$0.04 & 6.37$\pm$0.07 & 4.93$\pm$0.07 & 3.20$\pm$0.05 & 4.05$\pm$0.10 & 5.65$\pm$0.05 & 6.34$\pm$0.03\\
  NGC4815 & 8.73$\pm$0.05 & 7.53$\pm$0.06 & 7.39$\pm$0.09 & 6.34$\pm$0.11 & 4.85$\pm$0.03 & 3.07$\pm$0.06 & 3.87$\pm$0.03 & 5.50$\pm$0.01 & 6.23$\pm$0.11\\
  NGC6633 &  -                       & 7.58$\pm$0.03 & 7.37$\pm$0.05 & 6.31$\pm$0.05 & 4.87$\pm$0.06 & 3.05$\pm$0.04 & 3.92$\pm$0.08 & 5.61$\pm$0.06 & 6.10$\pm$0.05\\
  NGC6802 & 8.74$\pm$0.09 & 7.69$\pm$0.05 & 7.53$\pm$0.04 & 6.36$\pm$0.06 & 4.92$\pm$0.03 & 3.23$\pm$0.07 & 3.99$\pm$0.02 & 5.65$\pm$0.04 & 6.24$\pm$0.05\\
  Be81         & 8.95$\pm$0.13 & 7.87$\pm$0.06 & 7.62$\pm$0.06 & 6.52$\pm$0.05 & 5.10$\pm$0.08 & 3.39$\pm$0.05 & 4.25$\pm$0.09 & 5.84$\pm$0.07 & 6.53$\pm$0.09\\
  Tr23          & 8.84$\pm$0.07 & 7.87$\pm$0.07 & 7.66$\pm$0.05 & 6.42$\pm$0.07 & 4.96$\pm$0.07 & 3.27$\pm$0.06 & 4.09$\pm$0.06 & 5.72$\pm$0.07 & 6.35$\pm$0.06\\
  NGC6005 & 8.85$\pm$0.03 & 7.82$\pm$0.02 & 7.64$\pm$0.03 & 6.46$\pm$0.03 & 5.02$\pm$0.03 & 3.29$\pm$0.04 & 4.13$\pm$0.03 & 5.75$\pm$0.04 & 6.39$\pm$0.03\\
  Pis18        & 8.74$\pm$0.02 & 7.69$\pm$0.02 & 7.54$\pm$0.01 & 6.33$\pm$0.07 & 4.89$\pm$0.20 & 3.19$\pm$0.04 & 4.00$\pm$0.05 & 5.61$\pm$0.05 & 6.22$\pm$0.20\\
  Tr20    	   &  -                      & 7.71$\pm$0.04 & 7.55$\pm$0.06 & 6.39$\pm$0.03 & 4.97$\pm$0.03 & 3.21$\pm$0.06 & 4.03$\pm$0.05 & 5.68$\pm$0.04 & 6.30$\pm$0.05\\
  Be44         & 8.84$\pm$0.20 & 7.91$\pm$0.01 & 7.73$\pm$0.02 & 6.49$\pm$0.08 & 5.13$\pm$0.03 & 3.34$\pm$0.07 & 4.24$\pm$0.04 & 5.97$\pm$0.03 & 6.45$\pm$0.03\\
  NGC2243 & 8.47$\pm$0.08 & 7.28$\pm$0.04 & 7.09$\pm$0.06 & 5.92$\pm$0.04 & 4.52$\pm$0.06 & 2.87$\pm$0.05 & 3.51$\pm$0.08 & 5.11$\pm$0.07 & 5.80$\pm$0.05\\
  Be25         & 8.90$\pm$0.18 & 7.44$\pm$0.12 & 7.26$\pm$0.08 & 6.04$\pm$0.11 & 4.69$\pm$0.08 & 3.05$\pm$0.09 & 3.70$\pm$0.07 & 5.28$\pm$0.08 & 5.96$\pm$0.08\\
\hline \hline
\end{tabular}
\end{center}
\label{tab_clu_abu}
\end{table*}

\begin{figure*}
   \centering
  \includegraphics[angle=90,width=1.05 \textwidth]{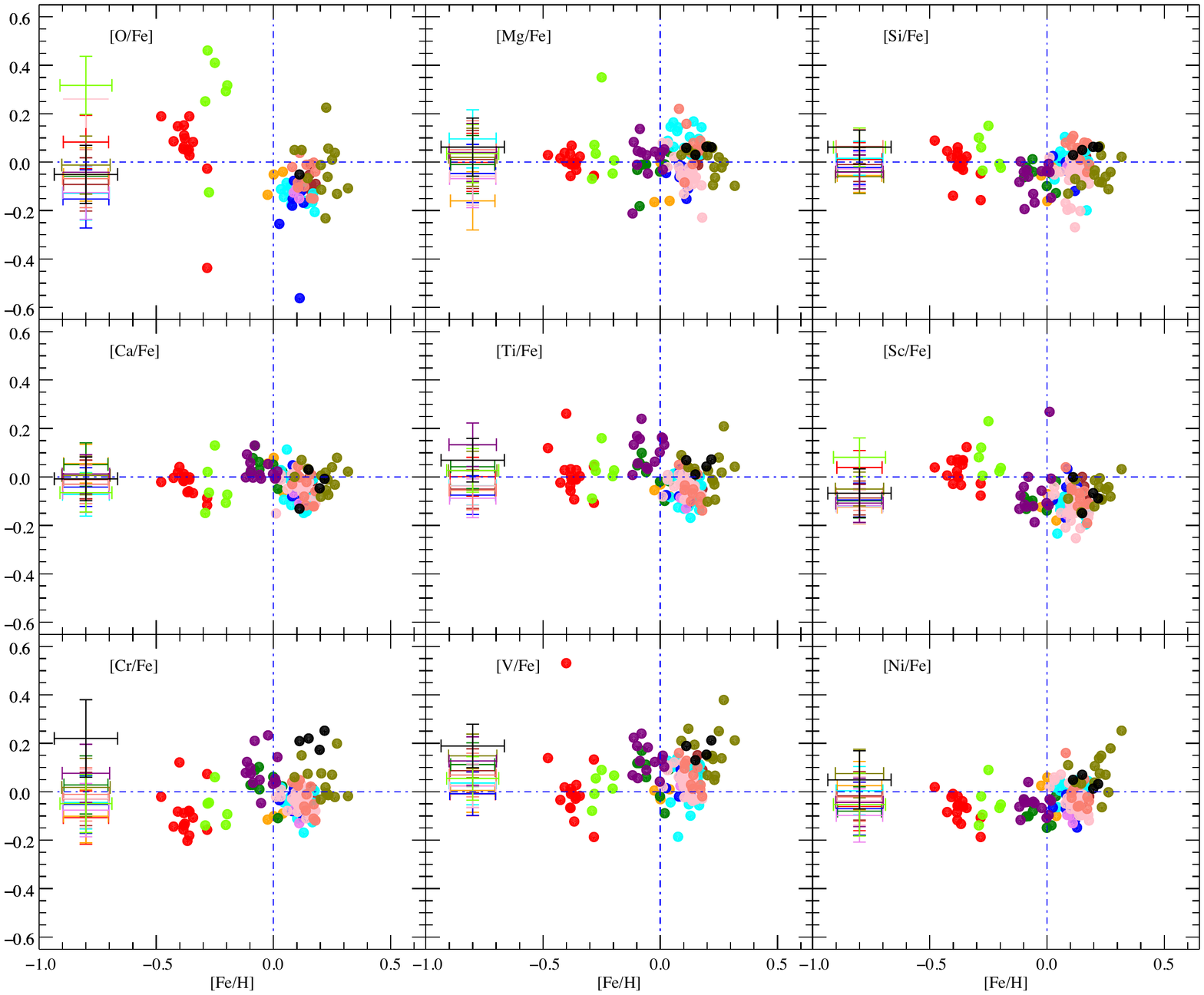}
    \caption{Abundance ratios of the member stars in our sample clusters. Colour code is the following:   NGC2243 red, 
NGC4815	 orange, 
NGC6005	brown, 
NGC6633	 green,  
NGC6705	 cyan, 
NGC6802	 blue,
Pismis18	 violet,    
Trumpler20 pink,  
Trumpler23 salmon,   
Be81		olive,
NGC2516	purple, 
Be25		light green,
Be44		   black. 
    The crosses on the left side of each panel represents the 1-$\sigma$ dispersion of the [Fe/H] and of the abundance ratios. 
  }
        \label{all_clusters}
 \end{figure*}

In Table~\ref{tab_clu_abufe} for each cluster we present the median abundance ratios with their 1-$\sigma$ dispersion normalised to the Solar scale in 
Table~\ref{tab_sun}.  We obtained them by 
computing the median values of the individual [X/Fe] in all the selected cluster member stars. 
These may slightly differ from the simple subtraction of the median [X/H] and [Fe/H].  
In Figure~\ref{all_clusters} the abundance ratios of the sample clusters are shown in the [X/Fe] vs [Fe/H] planes. Individual member stars are shown, together with the 1-$\sigma$ dispersion.  We note that for most clusters and elements there are small internal dispersions. 
However for some elements, as for instance oxygen, the dispersion is higher because of intrinsic difficulties in measuring them.  
\begin{table*}
\begin{center}
\caption{Clusters' abundance ratios}
\tiny
\begin{tabular}{lrrrrrrrrr}
\hline \hline
  \multicolumn{1}{c}{Id} &
  \multicolumn{1}{c}{[OI/Fe]} &
  \multicolumn{1}{c}{[MgI/Fe]} &
  \multicolumn{1}{c}{[SiI/Fe]} &
  \multicolumn{1}{c}{[CaI/Fe]} &
  \multicolumn{1}{c}{[TiI/Fe]} &
  \multicolumn{1}{c}{[ScII/Fe]} &
  \multicolumn{1}{c}{[VI/Fe]} &
  \multicolumn{1}{c}{[CrI/Fe]} &
  \multicolumn{1}{c}{[NiI/Fe]} \\
\hline
 NGC2516 &  -             		& 0.04$\pm$0.07 & -0.04$\pm$0.09 &  0.01$\pm$0.06 & 0.13$\pm$0.10 & -0.11$\pm$0.08 & 0.13$\pm$0.08 &  0.08$\pm$0.09 & -0.04$\pm$0.06\\
  NGC6705 & -0.13$\pm$0.07 	& 0.10$\pm$0.07 &  0.02$\pm$0.07 & -0.07$\pm$0.09 &-0.04$\pm$0.09 & -0.12$\pm$0.07 & 0.04$\pm$0.11 & -0.04$\pm$0.07 &  0.00$\pm$0.06\\
  NGC4815 & -0.05$\pm$0.05 	&-0.16$\pm$0.06 & -0.05$\pm$0.10 &  0.05$\pm$0.11 &-0.05$\pm$0.04 & -0.12$\pm$0.07 & 0.00$\pm$0.04 & -0.10$\pm$0.04 &  0.02$\pm$0.11\\
  NGC6633 &  -             		&-0.01$\pm$0.07 & -0.04$\pm$0.08 &  0.05$\pm$0.08 & 0.04$\pm$0.09 & -0.10$\pm$0.07 & 0.11$\pm$0.10 &  0.03$\pm$0.09 & -0.08$\pm$0.08\\
  NGC6802 & -0.15$\pm$0.09 	&-0.05$\pm$0.06 & -0.02$\pm$0.05 & -0.04$\pm$0.06 &-0.07$\pm$0.04 & -0.09$\pm$0.08 &-0.01$\pm$0.04 & -0.05$\pm$0.05 & -0.07$\pm$0.05\\
  Be81    & -0.01$\pm$0.13 	& 0.02$\pm$0.09 & -0.06$\pm$0.09 &  0.00$\pm$0.08 & 0.03$\pm$0.10 & -0.05$\pm$0.08 & 0.15$\pm$0.11 &  0.02$\pm$0.10 &  0.07$\pm$0.11\\
  Tr23    & -0.07$\pm$0.07 	& 0.05$\pm$0.08 &  0.06$\pm$0.06 & -0.03$\pm$0.07 &-0.05$\pm$0.07 & -0.07$\pm$0.07 & 0.07$\pm$0.07 & -0.01$\pm$0.07 & -0.02$\pm$0.07\\
  NGC6005 & -0.09$\pm$0.04 	& 0.01$\pm$0.02 & -0.01$\pm$0.03 & -0.01$\pm$0.03 &-0.05$\pm$0.03 & -0.09$\pm$0.05 & 0.09$\pm$0.03 & -0.03$\pm$0.04 & -0.02$\pm$0.03\\
  Pis18   & -0.13$\pm$0.03 	&-0.07$\pm$0.02 & -0.03$\pm$0.02 & -0.07$\pm$0.07 &-0.09$\pm$0.20 & -0.12$\pm$0.04 & 0.02$\pm$0.05 & -0.08$\pm$0.05 & -0.10$\pm$0.20\\
  Tr20    &  -  	   			&-0.06$\pm$0.05 & -0.04$\pm$0.07 & -0.03$\pm$0.05 &-0.04$\pm$0.05 & -0.12$\pm$0.07 & 0.03$\pm$0.06 & -0.03$\pm$0.05 & -0.04$\pm$0.07\\
  Be44    & -0.05$\pm$0.20 	& 0.06$\pm$0.05 &  0.06$\pm$0.05 & -0.00$\pm$0.09 & 0.07$\pm$0.06 & -0.07$\pm$0.08 & 0.19$\pm$0.06 &  0.22$\pm$0.06 &  0.05$\pm$0.06\\
  NGC2243 &  0.08$\pm$0.08 	& 0.00$\pm$0.06 &  0.01$\pm$0.07 & -0.01$\pm$0.06 & 0.00$\pm$0.07 &  0.04$\pm$0.07 &-0.01$\pm$0.09 & -0.11$\pm$0.08 & -0.06$\pm$0.07\\
  Be25    &  0.32$\pm$0.18 	& 0.03$\pm$0.13 &  0.06$\pm$0.09 & -0.06$\pm$0.12 & 0.03$\pm$0.10 &  0.08$\pm$0.10 & 0.05$\pm$0.08 & -0.05$\pm$0.09 & -0.05$\pm$0.09\\
 \hline \hline
\end{tabular}
\label{tab_clu_abufe}
\end{center}
\end{table*}

\section{The inner-disc giant and Solar neighbourhood dwarf samples}
\label{field}
To complement our cluster sample, we consider also  the full {\sc idr4} database of stellar parameters and abundances, extracting all field stars observed with UVES 580 belonging 
to the stars in the Milky Way sample, and in particular to the  Solar neighbourhood sample (\texttt{GES\_TYPE = }`GE\_MW') and to the inner disc sample  (\texttt{GES\_TYPE = }`GE\_MW\_BL').
For these stars, we compute ages and distances. 
Our method consists in a projection of the stellar parameters on a set of  isochrones \citep{bressan12}, thus obtaining a simultaneous determination of distance and estimation of age. 
The details of the method are described in \citet{kor11}, with the updates of \citet{rc14} and \citet{kordo15}. 
To compare with our cluster sample, we selected stars with age$<$5 Gyr. 
In addition we selected only stars with $|z|<$0.20 kpc, thus having a high probability to belong to the thin disc.   

Our final samples contain: 33 stars in the GE\_MW sample and 26 stars in the GE\_MW\_BL sample with ages $<$5 Gyr and belonging to the thin disc population.  Their stellar parameters, ages, distance, heights on the Galactic plane and abundances are presented in Appendix A, Tables~A.3 and A.4. 

\begin{figure}
   \centering
  \includegraphics[width=0.45\textwidth]{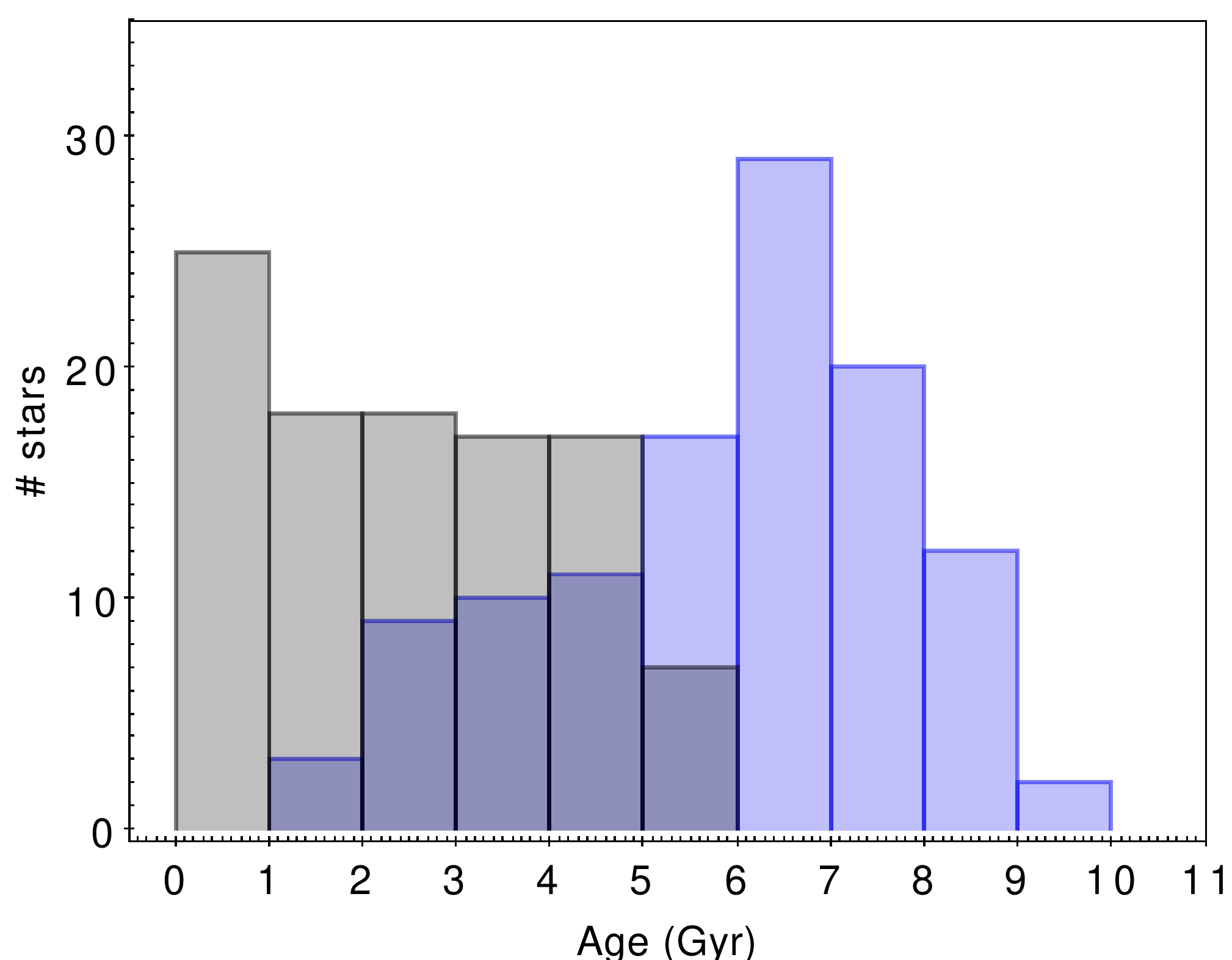}
    \caption{Histograms of the stellar ages in the GES Milky Way field sample in the  thin disc (in blue)  and in the open cluster sample (in grey).  
    }
 \label{histo}
 \end{figure}

These numbers have to be compared with 113 and 109 thin disc stars (defined on the basis of their height, z,  above the plane, $|z|<$0.20 kpc) of all ages in the GE\_MW  and GE\_MW\_BL samples, respectively. Thus only about 30\% of the thin disc stars 
in the GES {\sc idr4} are younger 
than 5~Gyr. If we make a more conservative selection, considering only stars younger than 2~Gyr, as most of the inner disc clusters, we have even lower numbers: 13 (11\%) and 7 (6\%) in 
the GE\_MW  and GE\_MW\_BL samples of thin disc stars, respectively. 
This highlights how young and intermediate age stars are poorly represented in the field populations and the  importance of clusters to characterise the recent abundance distribution 
in the thin disc. 
The histograms of the stellar ages of the  Milky Way field sample in  the thin disc and in the open cluster sample are shown in Figure~\ref{histo}. 
In the histogram all ages are consistently  computed with the projection  on  isochrones method, and that can slightly differ for the open cluster stars from the ages reported in Table~\ref{tab_clu_par}.

\section{The radial distribution of abundance ratios in clusters and field stars}
\label{radial}
Young and intermediate-age open clusters, together with the sample of thin disc field stars with age$<$5 Gyr--whose distances and especially ages are, however, much more uncertain-- represent a unique (and sometimes neglected)  
constraint to the shape of the abundance spatial distributions at recent epochs. 
A number of studies, including the recent results of the Apache Point Observatory Galactic Evolution Experiment (APOGEE) sample \citep{hayden14,hayden15}, 
have shown the spatial distributions of the abundances and abundance ratios of field stars (for instance, radial and vertical gradients and azimuthal variations). 
However, these studies are mainly based on field stars, representative of older populations and consequently they are affected by radial migration. 
Open clusters are a valuable alternative tool to study them, being on average younger, and therefore a better tracer of the gradients in the disc out of which the most recent stars formed, 
 as also shown in the recent APOGEE works on open cluster radial metallicity gradient \citep{frincha13, cunha16}.

\subsection{Abundance ratio gradients}

\begin{figure*}
   \centering
  \includegraphics[angle=90,width=0.95\textwidth]{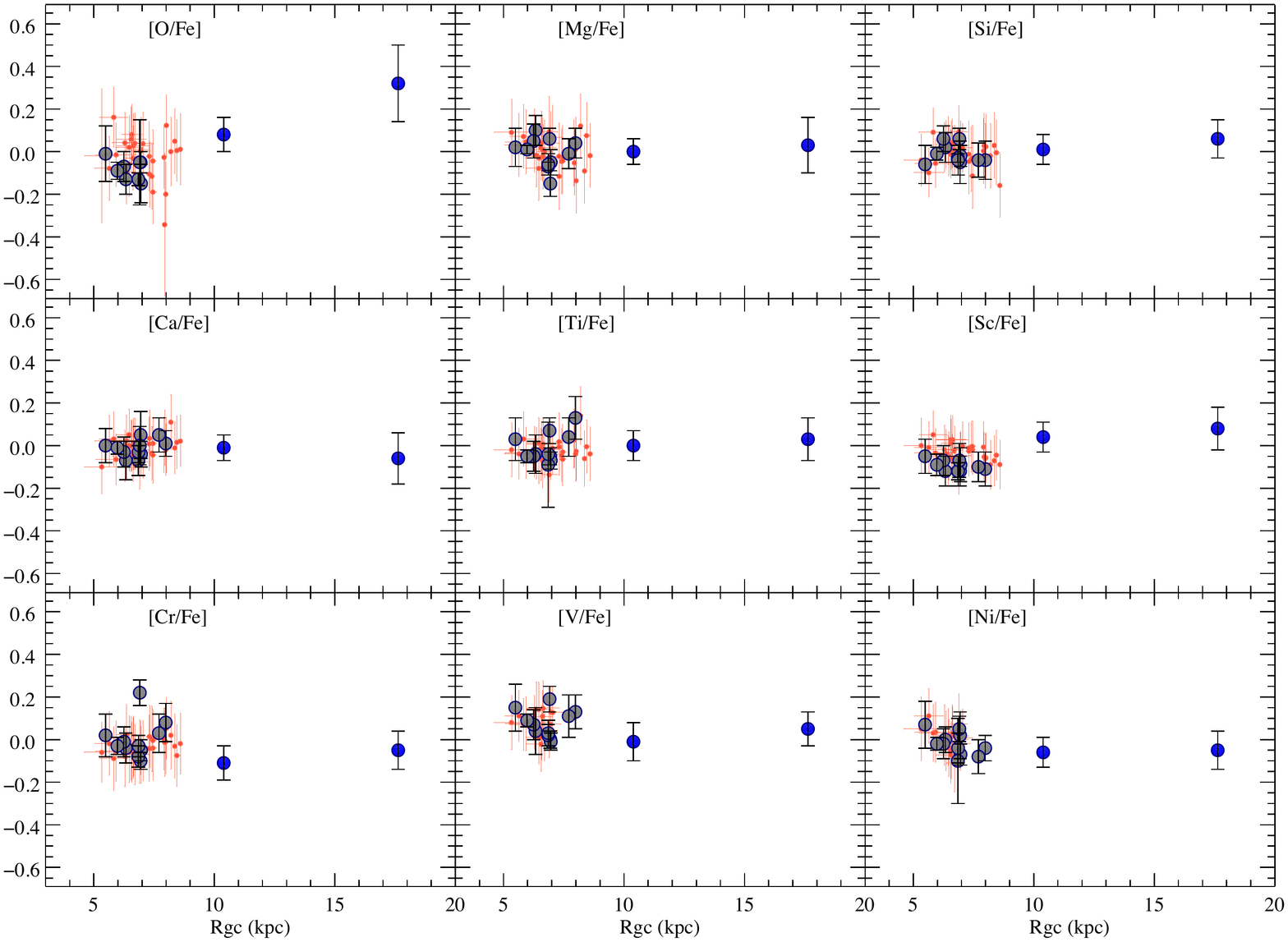}
    \caption{Radial abundance ratio gradients of elements belonging to different nucleosynthesis channels. The cluster median abundances are shown with larger circles, binned by 
    ages (in grey clusters with age$<$2 Gyr and in blue with age$>$2 Gyr). 
   The smaller red circles are the abundance ratios of the 'young' field stars in the thin disc with their errors.     }
        \label{grad_elfe_all}
 \end{figure*}

In Figure~\ref{grad_elfe_all}, we present for our cluster and field star samples the radial abundance ratio distributions of  the $\alpha$-elements [O/Fe],  [Mg/Fe], [Si/Fe], [Ca/Fe], [Ti/Fe] and of 
the iron peak-elements [Sc/Fe], [Cr/Fe], [V/Fe] and [Ni/Fe]. 
Plotting the abundance ratio over iron allow us to better appreciate the differences between the radial behaviour  of each element.  
The smaller circles in Figure~\ref{grad_elfe_all}  show the distribution of elemental abundances of thin disc field stars -- with  ages lower than 5~Gyr-- in the {\sc idr4} UVES sample.

The measurement of the oxygen abundance is based on a single absorption line at $\lambda$~6300 \AA, and thus it is quite 
difficult to obtain it, especially in dwarf stars.  For this reason, in Figure~\ref{grad_elfe_all} we do not plot [O/Fe] of dwarf field stars. 
Even considering only giant stars, the sample of field stars is indeed quite dispersed, whereas the open clusters seem to define an increasing trend towards the outer 
regions of the disc.  The inner-disc clusters have, on average, sub-Solar [O/Fe], while the two outer-disc clusters reach positive values of [O/Fe]. 

[Mg/Fe] is almost flat all over the disc, with a hint of increasing [Mg/Fe] in the inner disc. No suggestions of Mg-enhancement in the outer disc, nor of 
Mg-depletion in the inner disc are evident from our data. 
Si, Ca, and Ti have all similar behaviours: they reach null values in the inner disc and they are  enhanced (0.05-0.2~dex) in the outer disc. 
For the iron-peak elements, from Sc to Ni, we have that [Sc/Fe] has a slight increases in the outer-disc, similarly to the 
$\alpha$-elements, while Cr, V and Ni are   almost constant at Solar values across the whole disc. 

\section{Abundance patterns  of clusters and field stars}
\label{patterns}

\begin{figure*}
   \centering
  \includegraphics[angle=90,width=1.05\textwidth]{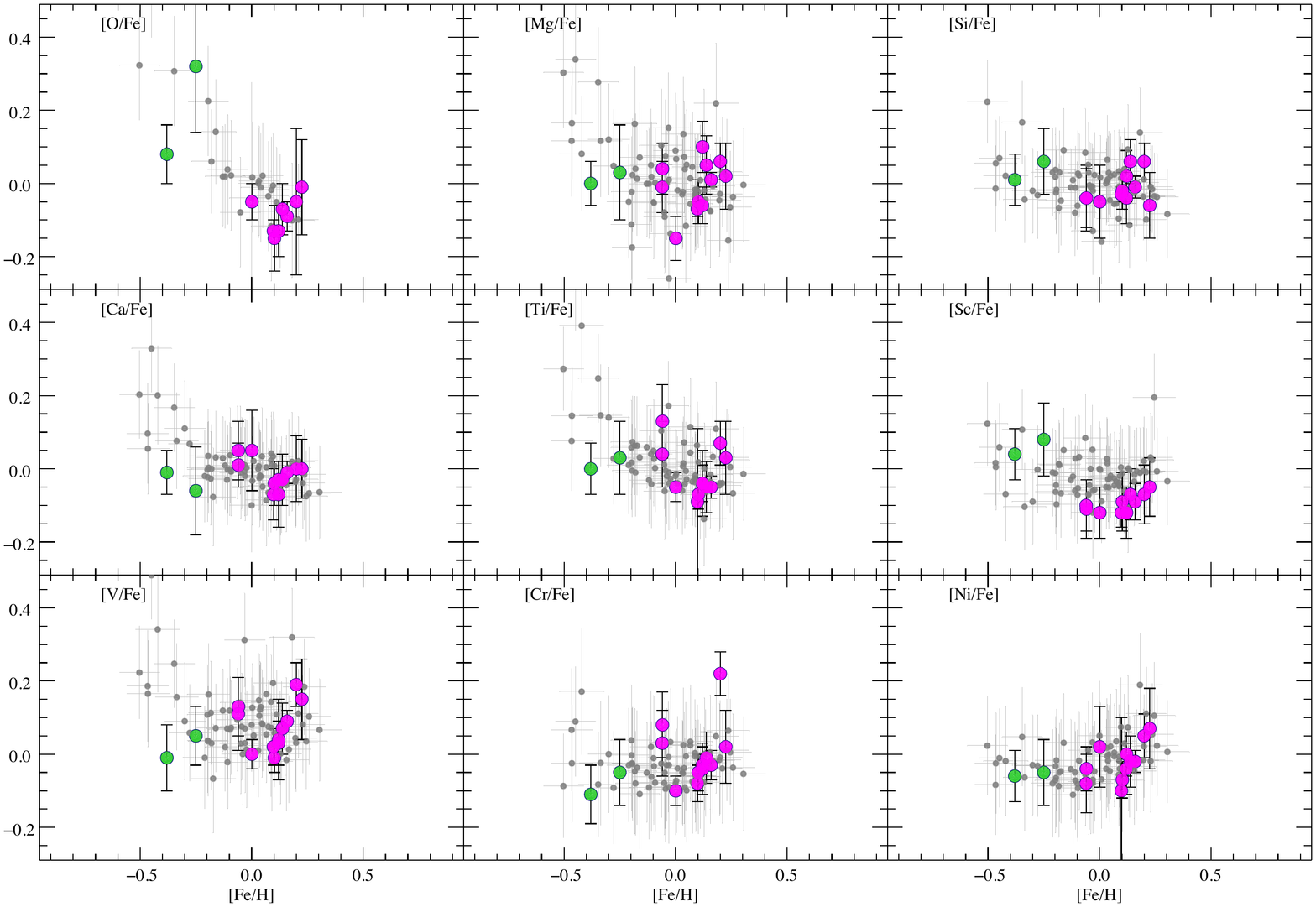}
    \caption{Abundance patterns of clusters (colour-coded  by distance, in magenta with R$_{\rm GC} < $8 kpc, and in green with with R$_{\rm GC} >$ 8 kpc) and of thin disc field stars (in grey).  }
        \label{fe_el_all}
 \end{figure*}

A classic alternative way to look at the Galactic chemical evolution is to consider the behaviour of the abundance ratios versus the iron abundance as a --non linear-- proxy of time. 
In Figure~\ref{fe_el_all}, we show the abundance patterns in the  [X/Fe] vs [Fe/H] planes of the cluster and thin disc population. 
For the thin disc, we include stars of all ages to explore also the low metallicity regime. 
However, the field sample is limited to the Solar neighbourhood by the GES selection function \citep[see][]{stonkute16}
and thus it does not reach much lower than [Fe/H]$<$-1~dex.  
The first five panels (from the left) show the abundance patterns of the $\alpha$-elements from oxygen to titanium. 
The observations show differences in the behaviour of the five elements: 
oxygen has the strongest trend, reaching negative values of [O/Fe] at super-Solar metallicities, and 
having positive values in the low metallicity regime. 
Magnesium in open clusters is essentially flat, while, at the lowest metallicities, the thin disc field stars reach 
positive values of [Mg/Fe]. In addition, contrary to oxygen, the trend of both open clusters and field stars with super-Solar metallicities indicates 
a [Mg/Fe] consistent with a slightly positive value. The behaviour of silicon 
is very similar to Mg, with a smaller dispersion in field stars with respect to Mg due to larger number of 
available lines for this element in the observed spectral range. [Si/Fe] is almost flat and 
slightly above zero 
in the super-Solar metallicity regime 0.0$<$[Fe/H]$<$0.5 dex.  However, there are some differences in the most metal poor regime sampled by our stars, where at [Fe/H]$\sim$-0.5~dex 
the few field stars reach higher [Mg/Fe] than [Si/Fe], and show a different behaviour with respect to the two outermost and most metal poor clusters. The differences might be related to 
the large errors on the determination of the field star ages (see Table A.1) that may lead us to assign them to an incorrect age bin. 
Calcium in field stars has a well-defined [Ca/Fe] enhancement towards the lowest metallicity, while it is almost flat in open clusters. 
Finally, titanium is very similar to calcium, having however a larger dispersion in both field and cluster stars abundances. 

The last four panels show the abundance ratio of some iron-peak elements. 
Scandium show differences between field star and open cluster abundances. 
If we consider field stars, [Sc/Fe] is flat across the metallicity range [-0.5,0.5], with the inner disc open clusters 
have a depleted [Sc/Fe] around -0.1~dex. 
V, Cr, and Ni have similar trends, being almost flat, with a slight enhancement 
in the super-Solar regime.

\section{The chemo-dynamical model}
\label{model}
We compare our observational results with the chemical evolution model of \citet{kubryk15a},
updated with recent sets of stellar yields for  stars of low and high masses (see below). The
model is described in details in \citet[][hereafter K15]{kubryk15a,kubryk15b}. In the following we
recall its main features.

The Galactic disc is gradually built up by infall of primordial gas in the potential well of a
{\em typical} dark matter halo with mass of 10$^{12}$M$_{\odot}$ whose evolution is obtained from
numerical simulations. The infall time-scales are shorter in the inner regions, while they
increase outwards reaching 7 Gyr at 7 kpc. The star formation rate depends on the local surface
density of molecular gas and is calculated with the prescriptions of \citet{br06}.

The model takes into account the radial flows of gas driven by a bar formed 6 Gyr ago which pushes
gas inwards and outwards of the corotation. Stars, but also clusters \citep[see, e.g.,][]{gust16} move radially due to  epicyclic motions
(blurring) and  variation in their guiding radius (churning), \citep[see e.g.][]{SB09}.
The innovative aspect of the model is  to account for the fact that radial migration moves around
not only {\em passive tracers} of chemical evolution (i.e. long-lived stars, keeping on
their photospheres the chemical composition of the gas at the time and place of their birth), but
also {\em active agents} of chemical evolution, i. e., long-lived nucleosynthesis sources
such as SNIa producing Fe and low mass stars producing s-process elements.

The K15 version of the model used for massive stars the metallicity-dependent yields from \citet{nomoto13},
while the version adopted in the present work  (K15-improved) uses the new metallicity-dependent
yields by Limongi \& Chieffi (in preparation) which include the effect of stellar rotation.

These yields are based on a new grid of massive stellar models that range in mass between 13 and 120 $\rm M_\odot$, 
initial Fe abundances [Fe/H]=0, -1, -2, -3 and initial equatorial rotational velocities $\rm v=0,~150,~300$ km~s$^{-1}$. 
The network adopted includes 335 isotopes (from neutrons to Bi$_{209}$) linked
by more than 3000 nuclear reactions. The initial composition adopted for the Solar metallicity models is the 
one provided by \cite{agss09}, which corresponds to a total metallicity $\rm Z=1.345\times10^{-2}$. 
For metallicities lower than Solar we assume a scaled Solar distribution for all the elements, 
with the exception of C, O, Mg, Si, S, Ar, Ca, and Ti which are assumed to be enhanced with 
respect to Fe. In particular we adopted [C/Fe]=0.18, [O/Fe]=0.47, [Mg/Fe]=0.27, [Si/Fe]=0.37, [S/Fe]=0.35, [Ar/Fe]=0.35, [Ca/Fe]=0.33, 
[Ti/Fe]=0.23 \citep{cayreletal04,spiteetal05}. As a consequence of these choices, the corresponding metallicity below Solar are 
$\rm Z=3.236\times10^{-3},~3.236\times10^{-4},~3.236\times10^{-5}$, respectively.
Stellar rotation has been included as described in detail in \cite{cl13} with the following
exceptions: {\em i)} an improved treatment of the angular momentum transport in the envelope of the stars and
{\em ii)} a detailed computation of the angular momentum loss due to stellar wind. At variance with \cite{cl13},
the efficiency of the rotationally driven mixing has been calibrated by requiring the fit to the observed nitrogen abundance 
as a function of the projected rotation velocity in the Large Magellanic Cloud samples of the FLAMES survey \citep{untetal09}.
The explosive nucleosynthesis has been computed in the framework of the kinetic bomb induced explosion 
by means of a PPM hydrocode, as described in \cite{cl13}. The kinetic energy injected to start the explosion has
 been calibrated to ejected 0.07$\rm M_\odot$ of $\rm ^{56}Ni$ for the models
ranging in mass between 13 and 25 $\rm M_\odot$. This choice leads to final kinetic energies of the ejecta in the range 
$\rm 1-3\times10^{51}~erg$ that are consistent with the average kinetic energy of a sample 
of core collapse supernovae reported by \citet{pp15,lyman16}. The explosion of the stars more massive 
than 25 $\rm M_\odot$ would require a kinetic energy significantly higher than $\rm 3\times10^{51}~erg$ and
we assume that these models fail to explode and collapse directly to a black hole. The yields of these stars contain therefore only the mass ejected through the wind. 

A phenomenological  rate of SNIa is adopted,  based on observations
of extragalactic SNIa, while their  yields are from \citet{iwamoto99}. The initial mass
function (IMF) of \citet{kroupa02}, with a slope 1.5 for the high masses, is used.
Finally, the formalism of single particle population is used to calculate the rate of ejecta (both
for stars and SNIa) as a function of time, because it can account for  the radial displacements of
nucleosynthesis sources and in particular of SNIa as discussed in \citet{kubryk13}.

\section{Comparison with the model and discussion}
\label{comparison}
\subsection{Radial metallicity gradient}

\begin{figure*}
   \centering
  \includegraphics[angle=90,width=0.95\textwidth]{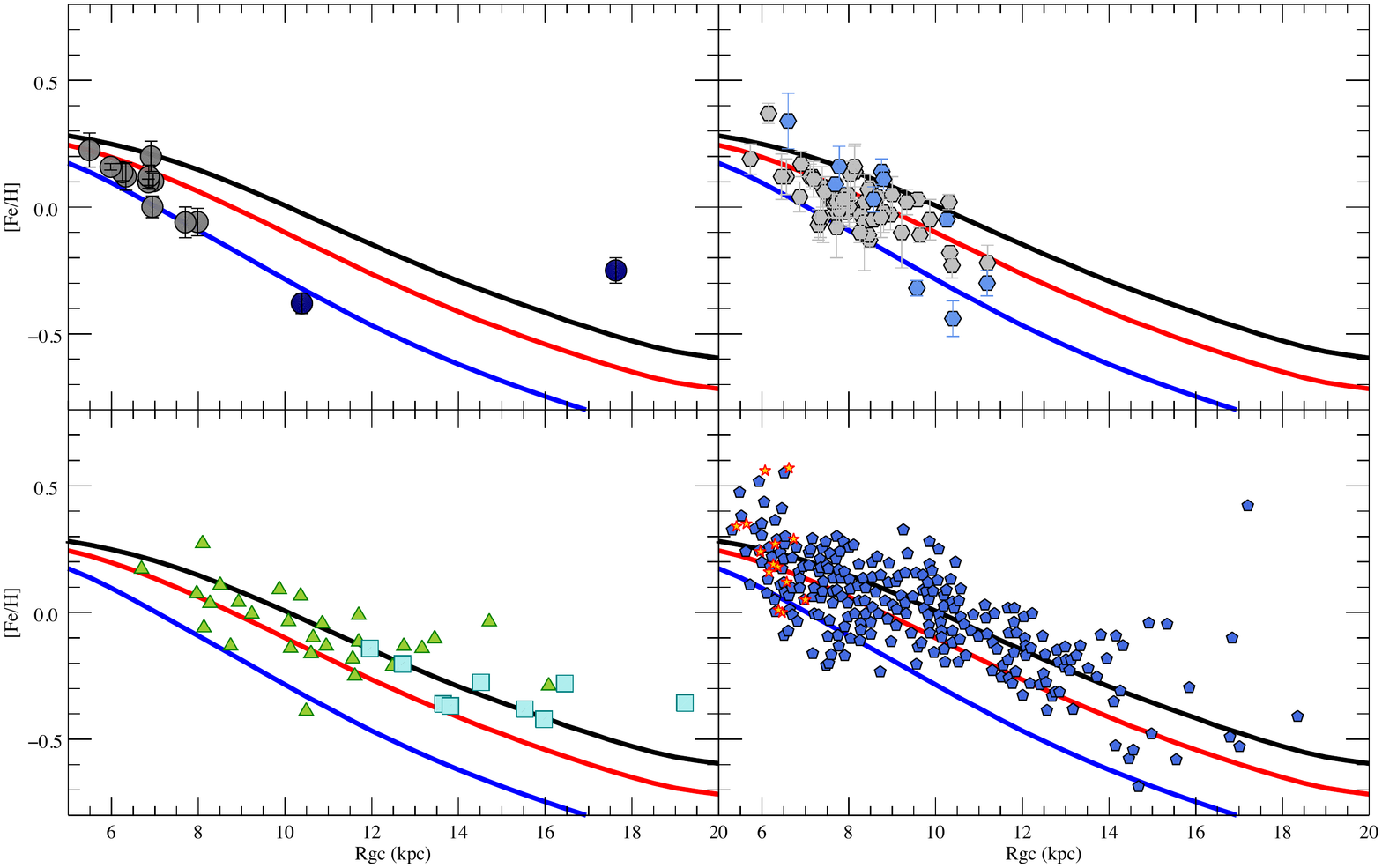}
    \caption{ Radial distributions of [Fe/H] for our open cluster sample (first panel-filled circles in grey the youngest clusters, age$<$2~Gyr, and in blue the oldest ones), for the \citet{netopil16}'s open clusters with high-resolution metallicities (second panel-filled hexagons in grey the youngest clusters, age$<$2~Gyr, and in blue the oldest ones), for the APOGEE \citep{cunha16} and \citet{CG16}'s open clusters (third panel, green triangles and cyan squares, resepctively), and for the Cepehids  \citep[fourth panel, stars and pentagons][respectively]{martin15, genovali14, genovali15}. The black curves represent the gradients of the elements over Fe in the  K15-improved model at the present time --updated with the new yields of the present work-- while the red and blue ones corresponds 
   to 2~Gyr and 5~Gyr ago, respectively.  }
        \label{grad_fe_model}
 \end{figure*}

In Figure~\ref{grad_fe_model} we show the radial distribution of [Fe/H] of  our sample open clusters, colour coded by age: younger than 2~Gyr and older than 2~Gyr.
We compare our results with some meaningful samples:  the literature compilation of \citet{netopil16} (second panel), selecting only open clusters with high resolution observations and determinations  of metallicity uncertainties. They are binned  in two age bins, similarly to  the GES sample: clusters younger than 2~Gyr, and clusters with 2~Gyr$<$Age$<$5~Gyr.  
In the third panel, we show two other literature open clusters samples: the APOGEE one from  \citet{cunha16} and the outer-disc clusters' one from \citet{CG16}.  
Finally, in the fourth panel we present the iron abundance of Cepheids \citep{martin15, genovali14, genovali15}.
We compare them with the  K15-improved model radial metallicity gradient at there different ages (present-time, 2~Gyr and 5~Gyr ago). 
Each sample, taken by itself, has its own limit, as for instance, 
limited statistics (first panel), combination of possible non homogeneous literature results (second panel), possibly poor membership and/or low number of stars per cluster (third panel), 
and finally large uncertainties on the metallicity determination of each Cepheid (fourth panel).
However, there is a general concordance in the shape of the radial metallicity gradients and in the metallicity reached in the four samples. 

The model curves are a good representation of the general radial behaviour of the [Fe/H] in the thin disc of our Galaxy, reproducing
 the declining gradient at least up to the optical radius of our Galaxy ($\sim$16~kpc) from which the open cluster samples show a 
departure from the negative gradient, reaching a plateau in metallicity \citep[e.g.][]{sestito08, m09, donati15, CG16, reddy16}. 
The plateau is not reproduced by the model and can be related to the high altitude of the
 outer-disc clusters above the plane, i.e., it is more properly the consequence of a vertical gradient than of a radial gradient and of radial migration and disc flaring \citep{minchev12}. 
 
In the first panel, we can see that our sample clusters located within the Solar circle (R$_{\rm GC} <$8~kpc) are all younger than 2~Gyr (the oldest one is Be~44 with an age of 1.6~Gyr), thus we do not expect that they moved so much from their birthplace.  
\citet{anders16} considered the possibility that already clusters with ages of about 2~Gyr, located from 5 to 8~kpc, from the Galactic Centre might be originated from regions located  more towards the centre. 
However, the metallicities of our clusters perfectly follow a radial decreasing gradient (see Figure~\ref{grad_fe_model}) with a small dispersion at each Galactocentric  radius. If the radial migration 
were the dominating process, we would expect a very scattered gradient and this is not the case. 
 
 Another notable result  that can be deduced from Figure~\ref{grad_fe_model} is the unexpected behaviour of the oldest clusters. 
 It is out of the scope of the present paper to discuss the time evolution of the radial metallicity gradient, however as already pointed out in several previous papers \citep[e.g.][]{jacobson16,anders16, gust16,spina17}  several old clusters (age$>$2~Gyr) are found to be more metal rich  than the younger clusters located at the same Galactocentric radius. 
 There are no old clusters in our sample in the inner disc to compare with the young and intermediate-age ones. 
 However, in the sample of \citet{netopil16} there is a super-position of the clusters with ages$>$2~Gyr and of the younger ones in the [Fe/H] versus R$_{\rm GC}$ plane, without 
 a clear time-evolution as indicated by the model. 
In a forthcoming paper (Kawata et al. in preparation),  we investigate the effect of radial migration on their location and metallicity. 

\subsection{Radial abundance ratio gradient}

\begin{figure*}
   \centering
  \includegraphics[angle=90,width=1.05\textwidth]{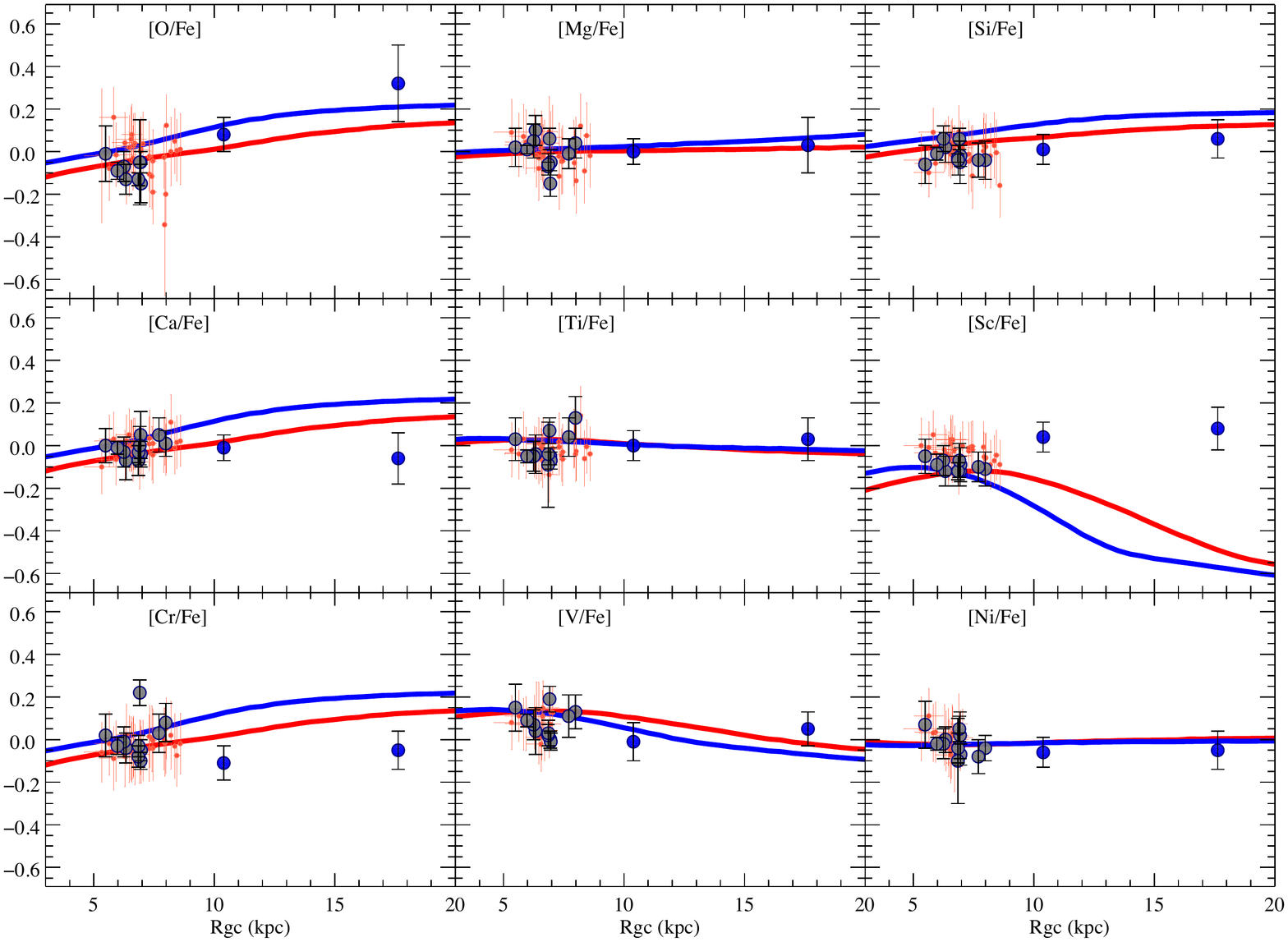}
    \caption{Radial abundance ratio gradients of elements belonging to different nucleosynthesis channels. For the observations, the symbols are 
    as in Figure~\ref{grad_elfe_all}.  The red curves represent  K15-improved model results at the present time, while the blue one corresponds 
   to 5~Gyr ago.     }
        \label{grad_elfe_all_model}
 \end{figure*}

In Figure~\ref{grad_elfe_all_model}, we present the abundance-ratio radial distributions of  the $\alpha$-elements [O/Fe],  [Mg/Fe], [Si/Fe], [Ca/Fe], [Ti/Fe] and of 
the iron peak-elements [Sc/Fe], [Cr/Fe], [V/Fe] and [Ni/Fe] for our clusters. We compare them with  the results of the  K15-improved model curves  at the  present time and at  a look-back time of 5 Gyr. 

Oxygen is mainly produced in the nucleosynthesis of massive stars (M$>$10 M$\odot$). These stars have short lifetimes  ($<$20 Myr) that do not give them enough time to migrate. 
Consequently, the radial O profile is not affected by radial migration, but, on the other hand, it is strongly impacted  by gas radial inflows. 
The presence of a bar that induces radial gas flows produces a non-monotonic gradient of [O/H] as a function of the Galactocentric radius.
In the parameterisation of \citet{kubryk15a}, in the inner regions (2-4~kpc)  the combination of  the bar and of  the metal-poor gas infall leads to a local depression of [O/H] with respect to the nearby
regions. On the other hand, the disc beyond 6 kpc is not affected by radial inflows, producing a decreasing gradient. 
In Figure~\ref{grad_elfe_all_model} (first row, first panel to the left)
we compare the predictions of the  K15-improved model  with our observations of clusters and field stars.  
The data of open clusters seem to support an [O/Fe] enhancement in the outer disc for the older clusters, 
while the conspicuous group of inner disc clusters presents a sub-Solar [O/Fe] as expected in the model predictions. It is  mainly driven by the different timescales for 
the formation of the inner and outer disc. The ratio between two elements generated by different kinds of stars is able to trace it. 

From an observational point of view, the radial distributions of [Mg/Fe], [Si/Fe], [Ca/Fe], and [Ti/Fe]  shown in Figure~\ref{grad_elfe_all} are all very similar, 
with a slightly increasing trend in the outer disc and Solar values in the inner disc.  
In the model we can distinguish between two kinds of behaviour: 
[Mg/Fe] and [Ti/Fe] are essentially flat and do not show any noticeable evolution with time, i.e. the radial gradients at the present time and 5~Gyr ago are almost similar, while
the model results for [Ca/Fe] and [Si/Fe] show a similar behaviour to [O/Fe] with differences between the curves at present time and 5~Gyr ago and an
enhancement in the outer disc. 
The $\alpha$ elements, from Si to the heavier elements Ca and Ti, are expected to have a non-negligible contribution from SNIa. 
The yields adopted in the K15 model, i.e. the \citet{iwamoto99}'s yields,  take the contribution of SNIa to the $\alpha$  elements into account. 
In addition, the new yields for massive stars used in the current version of the K15 model (Chieffi \& Limongi, in preparation) take into account the stellar rotation and the metallicity 
dependence. 

Comparing with the observations, we have that the model curve of [Mg/Fe] is in very good agreement with the observations of both clusters and field stars, 
and it is very different from the predictions of other chemical evolution models \citep[see, e.g.][]{minchev14}. 
For Si and Ca,  while the younger clusters (age$<$2~Gyr) are in good agreement with the model results, 
the two older and outer-disc clusters do not show the enhancement that is appreciable in [O/Fe]. 

One of the most important results of Figure~\ref{grad_elfe_all_model} is the nice agreement between the observed and modelled radial behaviour of [Mg/Fe]. 
This is indeed the first time, to our knowledge, that a chemo-dynamical model is able to distinguish between the evolution of [O/Fe] and [Mg/Fe]. 
O and Mg are considered to have both a predominant production in massive stars. However, they are produced during different burning phases in the evolution of massive stars:
oxygen is produced during the hydrostatic burning in the He-burning core and in the C-shell and it is expelled during the pre-supernova phase, in which the final yield can be slightly modified during the explosive Ne burning \citep[see, e.g.]{mmh}; magnesium is produced during the hydrostatic burning in the C shell and in the explosive burning of Ne, with a non negligible contribution of this latter process. These differences  can explain their abundance distributions.  

The radial trend of Mg is much more similar to that of Si, Ca and Ti,  and this is presumably an indication of common sites and processes of production. 
\citet{romano10} already noticed that the flattening in the [Si/Fe] vs. [Fe/H] plane traced by observations at super-Solar metallicities \citep{bensby05} requires 
a  source of Si enrichment during the latest phases of Galactic chemical evolution. They suggest that this source can be obtained, for instance,  from high-metallicity massive stars and/or SNIa. 

Introducing the new metallicity dependent yields of massive stars that induce a production of elements such as Mg at recent epochs in the Galaxy lifetime 
has for the first time reproduced the radial gradient of [Mg/Fe]  which is essentially flat. 
A similar conclusion was reached by \citet{romano10} who suggested the need for either a revision of current SNII and/or HN yields for Solar and/or higher than Solar metallicity 
stars, or larger contributions to Mg production from SNIa, or significant Mg synthesis in low- and intermediate-mass stars, or a combination of all these factors. 
This is in agreement with what was found by \citet{chiappini05} who stated that larger quantities of Mg (at least a factor of 10 more than current theoretical predictions of either 1-D or multi-D models) need to be produced 
in recent epochs, suggesting a production in SNIa.

The iron-peak group includes many elements ranging from Sc to Ge in the periodic table.
They are produced in different and complex nucleosynthesis processes
that result in a Galactic chemical evolution of their abundances not always following that of iron \citep[cf.][]{battistini15}. 
Here we consider the most representative elements of the iron-group available in our spectral range:  Sc, V, Cr, and Ni. 
SNIa contribute very significantly to the iron-peak elements, producing a  very little amount of elements lighter than
Al \citep[see, e.g.][]{iwamoto99, woosley09}. 
In addition to the component from SNIa,  Sc is also produced in the ejected layers of core-collapse SNe and that its  abundance is then further enhanced by neutrino-nucleus interactions (Yoshida et al. 2008).
Ni, V and Cr are also synthesised in massive stars \citep{WW95, LC03}.
We refer to \citet{romano10} for a complete description of the nucleosynthesis processes involved in the production of these elements. 

From the latest panels of Figure~\ref{grad_elfe_all_model}, we have good agreement for Cr, V, and Ni with the results of  K15-improved model. 
Cr and, to some extent, V abundances show a small systematic offset from the predictions,  while Sc is underproduced  by the model and the global 
trend traced by the open clusters is not followed. 
This indicates that the prescriptions of the model for the nucleosynthesis of Sc need to be updated. 

\subsection{Abundance patterns}

\begin{figure*}
   \centering
  \includegraphics[angle=90,width=1.05\textwidth]{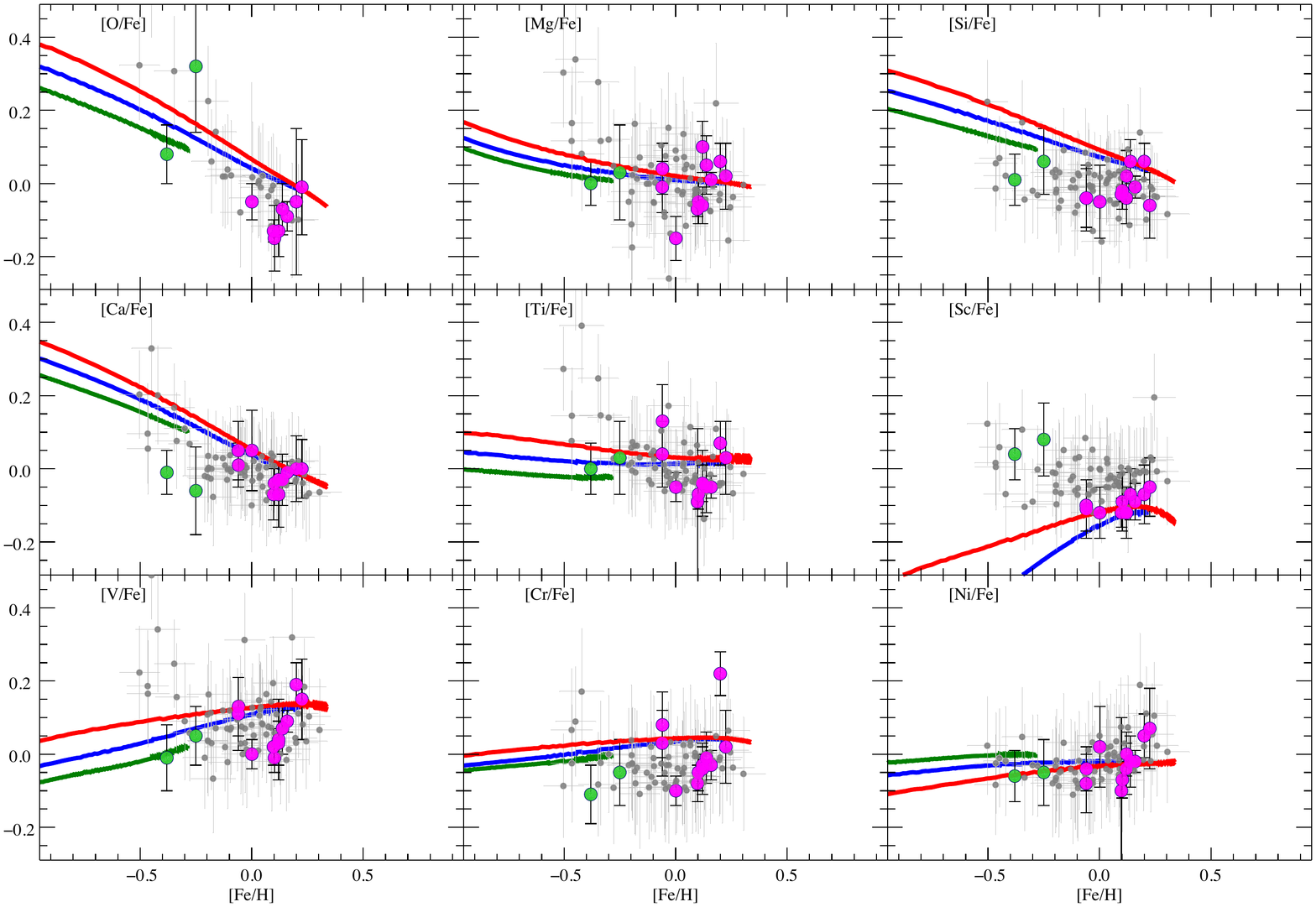}
    \caption{Abundance patterns of open clusters with the same symbols as in Fig.\ref{fe_el_all}, compared with the  curves of the K15-improved model.  The model curves are computed  for three Galactocentric radii: 6~kpc (red), 8~kpc (blue) and 15~kpc (green). 
}
        \label{fe_el_all_model}
 \end{figure*}

In Figure~\ref{fe_el_all_model} we present the abundance ratios as a function of the metallicity together with 
the results of the  K15-improved model for three Galactocentric radii: 6~kpc, 8~kpc and 15~kpc 
In the case of the $\alpha$-elements, the model predict two broad behaviours: oxygen, silicon, 
and calcium have a continuous decreasing trend 
up to super-Solar metallicities, reaching negative values for [O/Fe] and [Ca/Fe], while 
arriving only to zero for [Si/Fe]. 
On the other hand, [Mg/Fe] and [Ti/Fe] become almost flat at [Fe/H]$\sim$-0.5. This is caused by two effects: 
the metallicity dependent yields of massive stars and the contribution of SNIa to their production.
For Ti the production in SNIa is the dominant one, while for Mg the most important contribution is from SNII and the production at later epochs is increased by the 
metallicity dependent yields of progenitors of SNII. 
The last four panels show the abundance ratio of the most representative iron-peak elements together with the model predictions. 
[Sc/Fe] is clearly the worst case for which the model is not able to reproduce the data. 
[V/Fe], [Cr/Fe] and [Ni/Fe] have similar behaviours, which are, however, not exactly flat.
The model is able to reproduce the slight enhancement towards Solar/super-Solar metallicities that 
indicate the differences in their nucleosynthesis  with that of iron.  

There is a good agreement of the cluster abundance of the inner and outer disc with the corresponding curves of the model: 
for most elements the outer clusters agree well with outermost plotted curve, while the abundance ratios of the 
group of nine inner disc clusters are located within the two curves 
corresponding to 6 and 8~kpc. 
This can explain the differences that can be seen in Figure~\ref{grad_elfe_all_model} in the patterns of field stars and open clusters: while 
the field star sample is limited to the Solar neighbourhood, the open clusters are located in a larger Galactocentric range. 

\subsection{On the inside-out scenario for the Galactic thin disc}

The inside-out scenario for the formation of the thin disc predicts a higher star formation rate in the inner parts due to the concurrence 
of the higher infall rate and of the more effective star formation. 
The effect of the inside-out formation is appreciable by the presence of negative radial metallicity gradients in most disc galaxies. 
However, as an effect of different time scales of iron and $\alpha$-elements, we should expect a 'positive' gradient of [$\alpha$/Fe] in the disc whose entity and 
slope tell us about the differences in the time scales of the formation of the different regions.  
The presence of this possible enhancement has been debated for a long time, with contrasting results obtained from different stellar populations: 
open clusters \citep[see, e.g.][]{carraro04,  yong05, bragaglia08}, field stars \citep{carney05, bensby11}, and Cepheids \citep{yong06}. 
A discussion on this issue using open clusters can be found in \citet{yong12}. 

The problem of many literature works has been to use an `average' $\alpha$- enhancement which is based on different combinations of some of the five more commonly measured elements in the stellar atmospheres 
of cool stars. The point is, as shown in Figures~\ref{grad_elfe_all} and \ref{fe_el_all}, that these five elements do not share the same nucleosynthesis, and this is especially true at Solar/super-Solar metallicities, which is the characteristics metallicity of the thin disc and it represents the metallicity range spanned by the open cluster population.  
While it can be acceptable to consider the $\alpha$-elements equivalent to study the dichotomy between the thin and thick discs at low metallicity,  they widely diverge 
from [Fe/H]$\sim$-0.5~dex to super Solar [Fe/H] as appreciable from   Figures~\ref{grad_elfe_all} and \ref{fe_el_all}. 
This was already noticed in the seminal work of \citet{PT95} where those authors already warned about the differences in the nucleosynthesis of the $\alpha$ elements and the risk 
to mix them in a common [$\alpha$/Fe] value. 

\begin{figure}
   \centering
  \includegraphics[angle=90,width=0.52\textwidth]{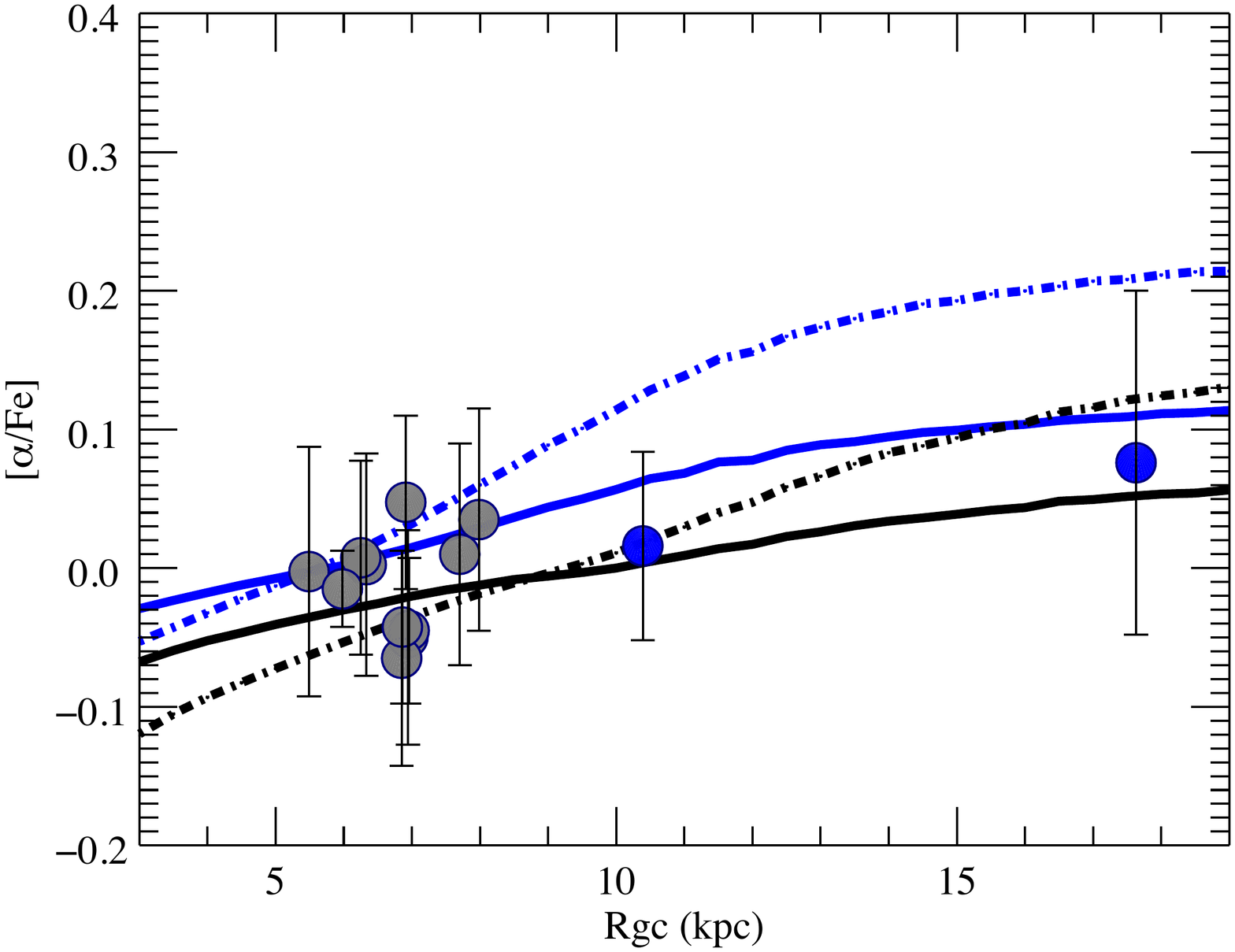}
    \caption{Global $\alpha$-enhancement: [$\alpha$/Fe]  vs. R$_{\rm GC}$ in the open cluster observations (colour coded by age as in Figure~6) and in the  K15-improved model (continuous lines, black at the  present time and blue 5~Gyr ago).  For comparison, the evolution of [O/Fe] vs. R$_{\rm GC}$ is shown (dashed-dotted lines, black at the present time and blue 5~Gyr ago). }
        \label{afeglobal}
 \end{figure}

In Figure~\ref{afeglobal} we show the `global' [$\alpha$/Fe], computed with O, Mg, Si, Ca, and Ti, as function of the  R$_{\rm GC}$ in the model and in the observations. The [$\alpha$/Fe]  both in model and observations have been computed in the same way, i.e. 
by computing the average of [X/Fe] (in some cases some elements are not available, thus the average has been computed with the remaining ones). This is the usual approximation adopted in the various literature studies. 
The observations are compared with the modelled  [$\alpha$/Fe] and with [O/Fe]: while for [O/Fe] we expect in the age interval spanned by open clusters to have in the 
outer parts of the Galaxy values ranging from [O/Fe]=0.1~dex to 0.2~dex, for  [$\alpha$/Fe]  the expected enhancement is much lower, from 0.05~dex to 0.1~dex. 
This due to the contribution to the average of elements such as Ti and Mg that behave as Fe at Solar and super-Solar metallicities. 
The combination of the five $\alpha$-elements produces an hybrid behaviour that can mask possibly expected differences 
between inner and outer disc populations. 

Thus, our final recommendation is to not use an average [$\alpha$/Fe] ratio at least for the typical metallicities of the thin disc, 
and we suggest to distinguish between the different channels of production of the different $\alpha$-elements when seeking subtle trends as the 
outer disc $\alpha$-enhancement or the inner disc $\alpha$-depletion.

\section{Summary}
\label{summary}

We analyse a sample of young and intermediate-age open clusters (age> 0.1~Gyr) in the fourth data release of the Gaia-ESO Survey. Using the recommended stellar parameters and elemental abundances of stars observed with UVES, we determine the median abundances of each cluster. We  determine statistical ages and distances of field stars observed with UVES
and we  select a sample of stars in the same age range of clusters. 
Using cluster and field star abundances, we derive the radial distribution of abundance ratios of several $\alpha$- and iron-peak elements, and their patterns as a function of metallicity, [Fe/H]. 
We notice important differences in the diverse classes of elements:  in particular we find that [O/Fe] has a different behaviour with respect to the other $\alpha$ elements, in particular Mg. 
We compare our observations, together with literature data, with the results  of chemical evolution models that include stellar migration and an updated set of stellar yields for massive stars. The model is able to reproduce the differences in the evolution of O and Mg, which are usually neglected but that  are important especially in the Solar and super-Solar metallicity regime. 

Finally we recommend to not use an average [$\alpha$/Fe] ratio at least for the typical metallicities of the thin disc. 
It is necessary to differentiate the channels of production of the $\alpha$-elements when searching for small trends as the 
inner disc $\alpha$-depletion and the 
outer disc $\alpha$-enhancement.

    \begin{acknowledgements}

The authors thanks the referee for her/his constructive report and Dr. Laura Inno for useful discussions. 
The results presented here benefited from discussions in three Gaia-ESO workshops supported by the ESF (European Science Foundation) through the GREAT (Gaia Research for European Astronomy Training) Research Network Program (Science meetings 3855, 4127 and 4415).
This work was partially supported by the Gaia Research for European Astronomy Training (GREAT-ITN) Marie Curie network, funded through the European Union Seventh Framework Programme [FP7/2007-2013] under grant agreement n. 264895. 
This work was partly supported support through the European Research Council grant 320360: The Gaia-ESO Milky Way Survey
G.T. and A.D. acknowledge support by the Research Council of Lithuania (MIP-082/2015).This research has been partially supported by the National Institute for Astrophysics (INAF) through the grant PRIN-2014 ("Transient Universe, unveiling new types of stellar explosions with PESSTOÓ).
F.J.E. acknowledges financial support from the Spacetec-CM project (S2013/ICE-2822).
S.F. and T.B. are supported by the project grant ÓThe New Milky WayÓ from the Knut and Alice Wallenberg Foundation.
Support for SD was provided by the ChileÕs Ministry of Economy, Development, and TourismÕs Millennium Science Initiative through grant IC120009, awarded to The Millennium Institute of Astrophysics, MAS. 
\end{acknowledgements}

\end{document}